\documentclass[aps,prl,twocolumn,superscriptaddress]{revtex4-1}
\usepackage{float}
\usepackage{amsmath,array,amssymb}
\usepackage{graphicx}
\usepackage{siunitx,bigints}
\usepackage{multirow}
\usepackage{amsfonts}

\usepackage{braket}

\usepackage{lineno}

    \DeclareMathOperator{\sech}{sech}
\usepackage{color}

\begin{document}

\title{Cascade Superfluorescence in Er:YLF
}%

\author{F. Chiossi}
\email[Electronic address: ]{federico.chiossi@unipd.it}

\author{C. Braggio}
\affiliation{Dip. di Fisica e Astronomia, University of Padova and INFN of Padova, I-35131 Padova, Italy}

\author{A. Khanbekyan}
\affiliation{Dip. di Fisica e Scienze della Terra, University of Ferrara and INFN of Ferrara, I-44122, Ferrara, Italy}

\author{G. Carugno}
\affiliation{Dip. di Fisica e Astronomia, University of Padova and INFN of Padova, I-35131 Padova, Italy}

\author{A. Ortolan}
\affiliation{INFN, Laboratori Nazionali di Legnaro, I-35020 Legnaro, Italy}

\author{G. Ruoso}
\affiliation{INFN, Laboratori Nazionali di Legnaro, I-35020 Legnaro, Italy}

\author{R. Calabrese}
\affiliation{Dip. di Fisica e Scienze della Terra, University of Ferrara and INFN of Ferrara, I-44122, Ferrara, Italy}

\author{A. Di Lieto}
\affiliation{Dip. di Fisica, University of Pisa, I-56127, Pisa, Italy}
\affiliation{NEST, Scuola Normale Superiore and Istituto Nanoscienze-CNR, I-56127 Pisa, Italy}

\author{L. Tomassetti}
\affiliation{Dip. di Fisica e Scienze della Terra, University of Ferrara and INFN of Ferrara, I-44122, Ferrara, Italy}

\author{M. Tonelli}
\affiliation{Dip. di Fisica, University of Pisa, I-56127, Pisa, Italy}
\affiliation{NEST, Scuola Normale Superiore and Istituto Nanoscienze-CNR, I-56127 Pisa, Italy}


\begin{abstract}
	
\noindent
We report the analysis of paired photon pulses arising from two cascading transitions in continuously pumped Erbium-doped YLiF$_4$ 1\% and 0.01\% crystals at 1.6\,K. The dependence of the pulse peak intensity on the squared number of involved Erbium ions, between 10$^{11}$ and 10$^{13}$, definitely identifies the cooperative nature of the two pulsed emissions, that are generated by the subsequent, spontaneous formation of coherent states. The observed fluctuations of the time interval between the paired pulses and, most importantly, its correlation with the second pulse duration, demonstrate that the Erbium ions coherence is indeed seeded by vacuum fluctuations.

\end{abstract}

\maketitle

\modulolinenumbers[5]

The coherent coupling of $N$ identical excited atoms results in a cooperative spontaneous emission in which the atomic transition rate is accelerated by a factor $N$\,\cite{Gross1982}. Rather than being prepared through external coherent sources\,\cite{Macfarlane2002,Klein2007}, the coupling can be seeded in uncorrelated identical emitters by their own independent spontaneous emission. Atomic coherence then exponentially spreads through the coherent emission that finally determines the atomic ensemble de-excitation. The resultant burst of coherent radiation is called superfluorescence (SF)\,\cite{Bonifacio1975}.

Starting from the first demonstration of SF from HF gas\,\cite{Skribanowitz1973} to the recent achievements in high pressure gases\,\cite{Mercadier2019}, nanostructured materials\,\cite{Raino2018} and colour centres in
diamond\,\cite{Angerer2018}, pulsed excitation has been employed to accomplish population inversion. A delayed directional emission has been searched as a signature of SF, as the maximal coherence is achieved after a random delay time $\tau_d$\,\cite{Vrehen1981,Mostowski1983a}. 
However, it is rarely considered that the pulsed excitation could determine the simultaneous formation of several, independent coherent subensembles. Since their related delay time is likely to be similar, the observed SF radiation burst could be a temporal superposition of few pulses. The result is a deviation from the expected sech-squared temporal profile, that introduces a systematic error in the estimate of the effective cooperativity factor among the radiators. 
Besides the excessive size of the active medium\,\cite{Kumarakrishnan1998}, the formation of independent subensembles is due to the different transition energy of the emitters. For instance, the superposition of two or three sech-squared pulses, ascribable to three well defined vibrational mode frequencies, has been detected in CH$_3$F gas\,\cite{Rosenberger1981}. The effect has been invoked  also to explain the observed amplitude modulations in SF pulses, arising from the interference of emissions by different hyperfine levels in Cs\,\cite{Vrehen1977}, Rb\,\cite{Marek1979} and Na\,\cite{Marek1980} vapours.  Indeed, in inhomogeneously broadened systems, as is the case for solid-state materials, 
the degree of cooperativity can be hardly inferred from the complex SF temporal profile\,\cite{Florian1984}. 

The superposition of the coherent emissions can be avoided by cw pumping the population inversion since there is no temporal correlation in the formation of different coherent subensembles. Pure sech-squared pulses have been reported under this pumping condition in Er:Y$_2$SiO$_5$ (Er:YSO), and the radiated intensity is in agreement with that expected for a single macrocoherent state\,\cite{Braggio2020}.  

\begin{figure}[h!]
	\includegraphics[width=1\linewidth]{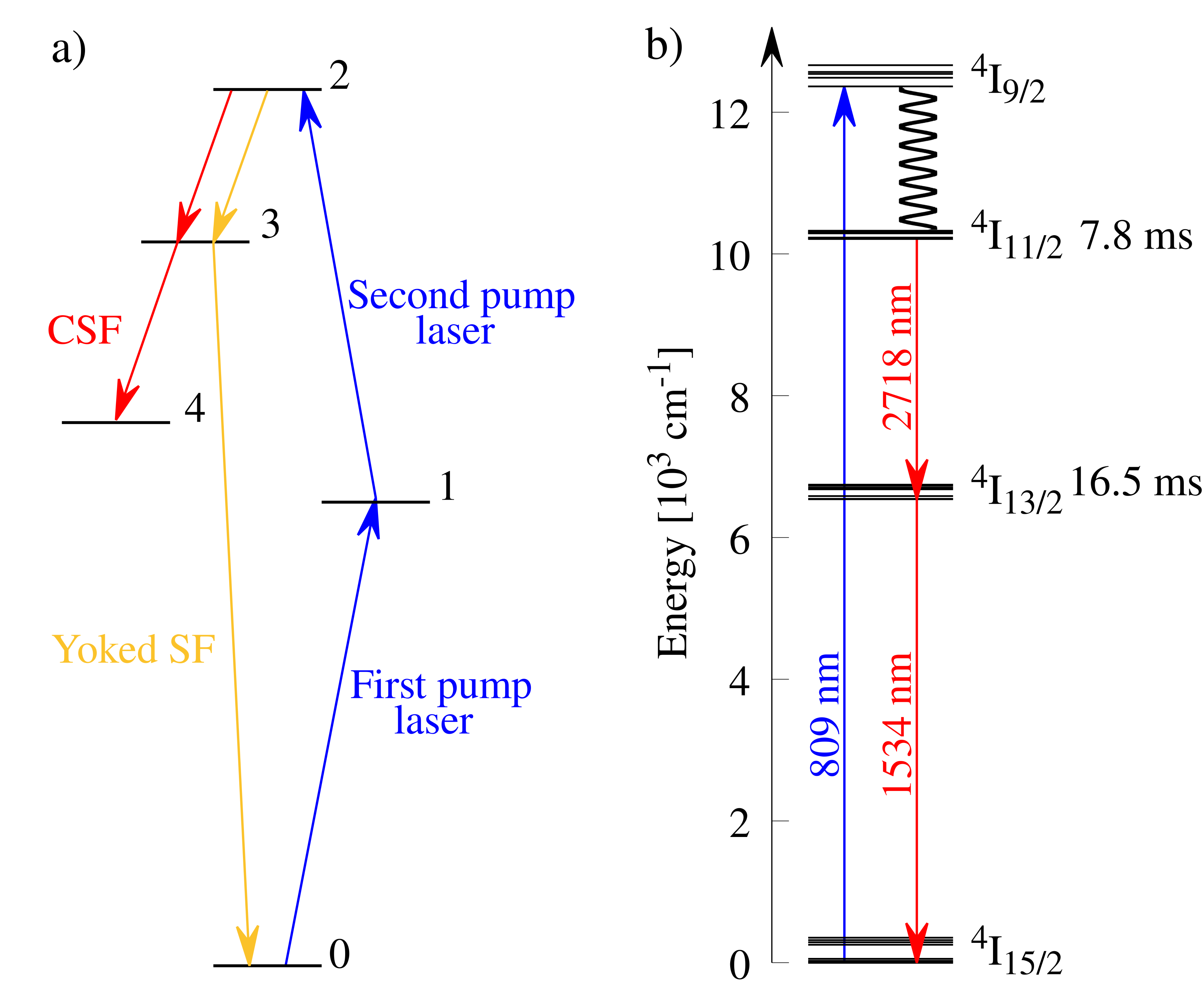}
	\caption{(a) General energy level and pumping schemes for the studies of cascade and yoked superfluorescence in gaseous systems. The two coherent emissions in CSF evolve independently and the SF pulses are temporally separated. This is not the case in yoked SF, where the second $3\rightarrow0$ transition is induced by the combination of the two coherences $0\rightarrow2$ and $2\rightarrow3$. (b) Energy level scheme of Erbium ions in YLF. The blue arrow indicates the cw laser excitation, the black wavy line the multiphonon relaxation, and the two red arrow the cascading superfluorescent transitions. The measured lifetime of the long-lived levels is also reported.
	}
	\label{Fig:1}
\end{figure}

In this work we study paired SF pulses emitted by Er:YLiF$_4$ (Er:YLF) crystals doped at 1\% and 0.01\% in a cascading transitions scheme. This phenomenon is referred to as cascade superfluorescence (CSF) as the atoms, driven to an intermediate level by a first SF process $2\rightarrow3$ (see Fig.\,\ref{Fig:1}), sequentially develop a second SF $3\rightarrow4$ transition. CSF was attained only in a scant number of gaseous systems\,\cite{Gross1976,Marek1979,Brechignac1981,Becker1991,Brownell1995} due to the demanding condition on the preparation of the excited state. If a coherence degree is initially induced between 0 and 2 levels, the onset of superfluorescence transition $2\rightarrow3$ is accompanied by the $3\rightarrow0$ coherent emission, called yoked superfluorescence, thereby suppressing the $2\rightarrow3\rightarrow4$ CSF\,\cite{Brownell1995,Paradis2008}.

An advantage of solid-state systems is the possibility to exploit multiphonon relaxation to accomplish a population inversion without initial coherence. However, the same process could be also responsible for the decrease of the steady-state population, which is instead essential to initiate SF in cw excitation. Actually, the SF regime requires that the characteristic superfluorescent time $\tau_R$, which scales with $N^{-1}$, must be much shorter than the atomic coherence time $T_2$. We thus select an active material with low phonon energy host matrix\,\cite{Zhang1994} in order to reduce the multiphonon relaxation rate\,\cite{Auzel1976}. In addition, the sample is maintained at 1.6\,K inside a superfluid Helium cryostat, described in Ref.\,\onlinecite{Braggio}, to ensure a long coherence time\,\cite{Minnegaliev2017,Kukharchyk2018}. Its three optical windows allow for laser pumping, investigating pulses emitted in the forward direction, along with monitoring the isotropic incoherent emission\,\cite{Braggio2020}.

\begin{figure}[h!]
	\includegraphics[width=1\linewidth]{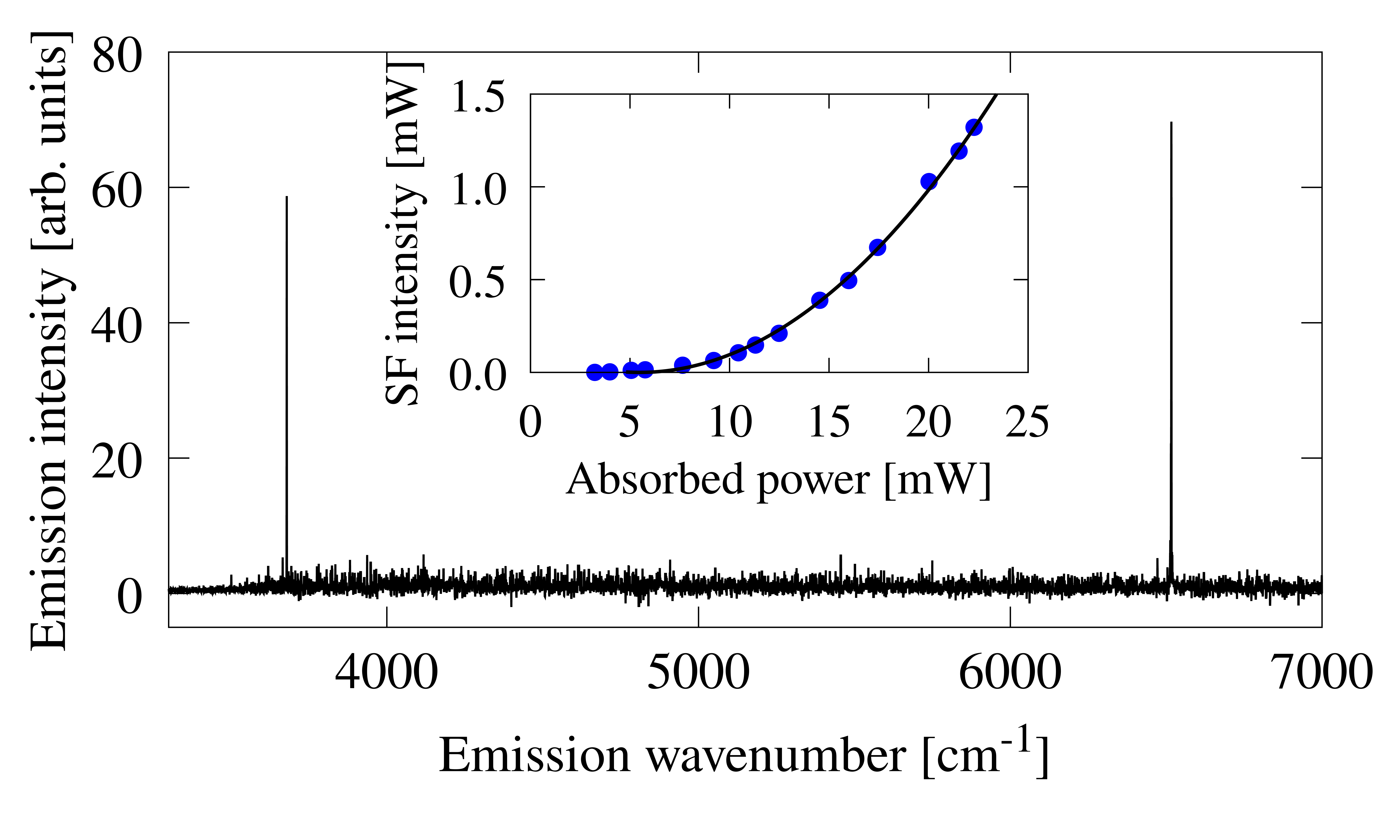}
	\caption{Spectrum of the forward emission recorded at a FTIR interferometer (Bruker, Equinox55) equipped with an InAs photodiode. The CSF emissions at 6516.7\,cm$^{-1}$ (1534.5\,nm) and 3678.8\,cm$^{-1}$ (2718.3\,nm) are identified. In the inset, the SF intensity at 1534\,nm measured with a Germanium sensor is plotted as a function of the absorbed laser power, proportional to the inverted population density in the Er:YLF 0.01\% crystal. Above the pumping threshold, data are fitted by a quadratic function (black line). }
\label{Fig:spet}
\end{figure}

\begin{figure}[!]
	\includegraphics[width=0.5\textwidth]{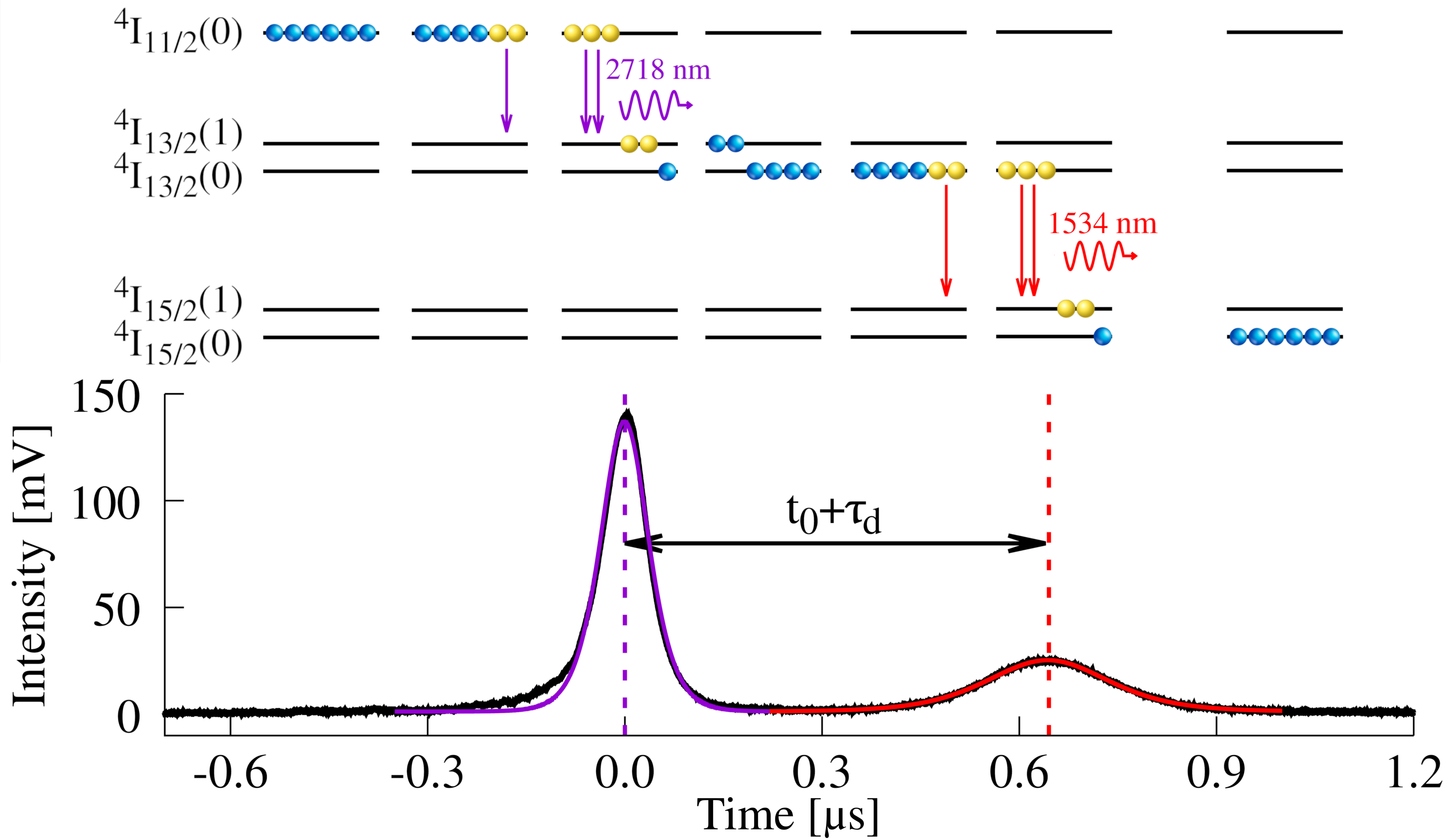}
	\caption{($Bottom$) Representative paired pulses recorded with a InAs detector (black line). Each pulse profile is fitted with the function $f(t) = P\sech^2((t-t_a)/2\tau_{\mbox{\tiny R}})$ (violet and red line) centered at time $t_a$, with maximum value $P$. ($Top$) Pictorial representation of cascade superfluorescence in Er$^{3+}$. Starting from uncorrelated  excited ions (blue spheres) in $^{4}\mathrm{I}_{11/2}(0)$ level, atomic coherence (yellow spheres) develops between the latter and the $^{4}\mathrm{I}_{13/2}(1)$ level, achieving its maximal degree at the peak of the 2718\,nm pulse. These ions then quickly thermalize to the long-lived $^{4}\mathrm{I}_{13/2}(0)$ level, where coherence is spontaneously established again, seeding SF emission to $^{4}\mathrm{I}_{15/2}(1)$ at 1534\,nm.}
	\label{Fig:2}
\end{figure}

Population inversion is accomplished on both the ms-long-lived levels $^{4}\mathrm{I}_{11/2}(0)$ and 	$^{4}\mathrm{I}_{13/2}(0)$ by tuning the wavelength of a cw Ti:Sapphire laser (10\,MHz linewidth) to the transition $^{4}\mathrm{I}_{15/2}(0) {}\rightarrow{}^{4}\mathrm{I}_{9/2}(0)$, as shown in Fig.\,\ref{Fig:spet}. Above a pumping threshold, we observe along the pump laser propagation direction pulsed emissions at 2718\,nm and 1534\,nm wavelength, whose time-average intensity scales superlinearly with the emitting level population (Fig.\,\ref{Fig:2}). The observed emissions correspond to the transitions ${}^4\mathrm{I}_{11/2}(0) \rightarrow 	{}^4\mathrm{I}_{13/2}(1)$ and ${}^4\mathrm{I}_{13/2}(0) \rightarrow {}^4\mathrm{I}_{15/2}(1)$, respectively\,\cite{Popova2000}.

These cascading transitions give rise to paired pulses as shown in Fig.\,\ref{Fig:2}. As the first pulse triggers the acquisition in the time domain, a stochastic distribution of delays of the following pulse at 1534\,nm is recorded. 
In our measurements, the upper level superfluoresces to a level that does not coincide with the starting level of the second SF. The time interval $\bar{\tau}_d$ between the cascade pulses therefore includes a time $t_0$, proportional to the non-radiative ${}^4\mathrm{I}_{13/2}(1) \rightarrow {}^4\mathrm{I}_{13/2}(0)$ transition time, in addition to the usual delay time $\tau_d$ of the second SF.

A competitive process that shares some features with SF is the amplified spontaneous emissions (ASE). In particular, the observation of directional, sech-squared pulses, along with a superlinear dependence of the time-average intensity with respect to the level population is not sufficient to discriminate SF from ASE. The latter dominates over SF for increasing temperatures\,\cite{Malcuit1987}, and SF could hardly be claimed in Er:YLF for temperatures up to 60\,K\,\cite{Hubert1990}. The definitive proof of cooperative emission is given by the $N^2$ dependence of the pulse peak intensity $R_p$.
As detailed in Ref.\,\cite{Braggio2020}, when the population inversion density is high, the single pass gain that the SF pulse experiences in its propagation through the medium must also be considered. The observed  $\bar{N}=N+N_0$ pulse photons can therefore be traced back to $N$ coherent atoms and $N_0$ uncorrelated atoms driven via stimulated emission. The scaling laws identifying SF process can thus be rewritten as functions of the observables $\bar{N}$, $\tau_R$ and $R_p$ in the forms\,\cite{Braggio2020}:
\begin{equation}
R_p=0.25\mu A \bar{N}(\bar{N}-N_0)
\label{eq:Rp1}
\end{equation}
\begin{equation}
\tau_{\mbox{\tiny R}}=\frac{1}{\mu A (\bar{N}-N_0)} = \frac{N_0}{8\,R_p}\left[1+\sqrt{1+\frac{16R_p}{\mu A N_0^2}}\right]
\label{eq:taup1}
\end{equation}
where $A$ is the spontaneous emission rate of the SF transition and $\mu$ is a factor that depends on the atomic sample geometry.

We have collected hundreds of paired pulses for both Er:YLF crystals using an InAs detector. In this case, the photon bunches are not fully collected and the SF equations are verified qualitatively for both the emission wavelengths by adding a scaling factor.

When the SF beam at 1.5\,$\mu$m is properly coupled to the sensitive area of an InGaAs photodiode, we are able to infer the coherent atom number $N$ by measuring $\bar{N}$ and by estimating $N_0$ with the fitting procedure.
\begin{figure}[!]
	\includegraphics[width=0.5\textwidth]{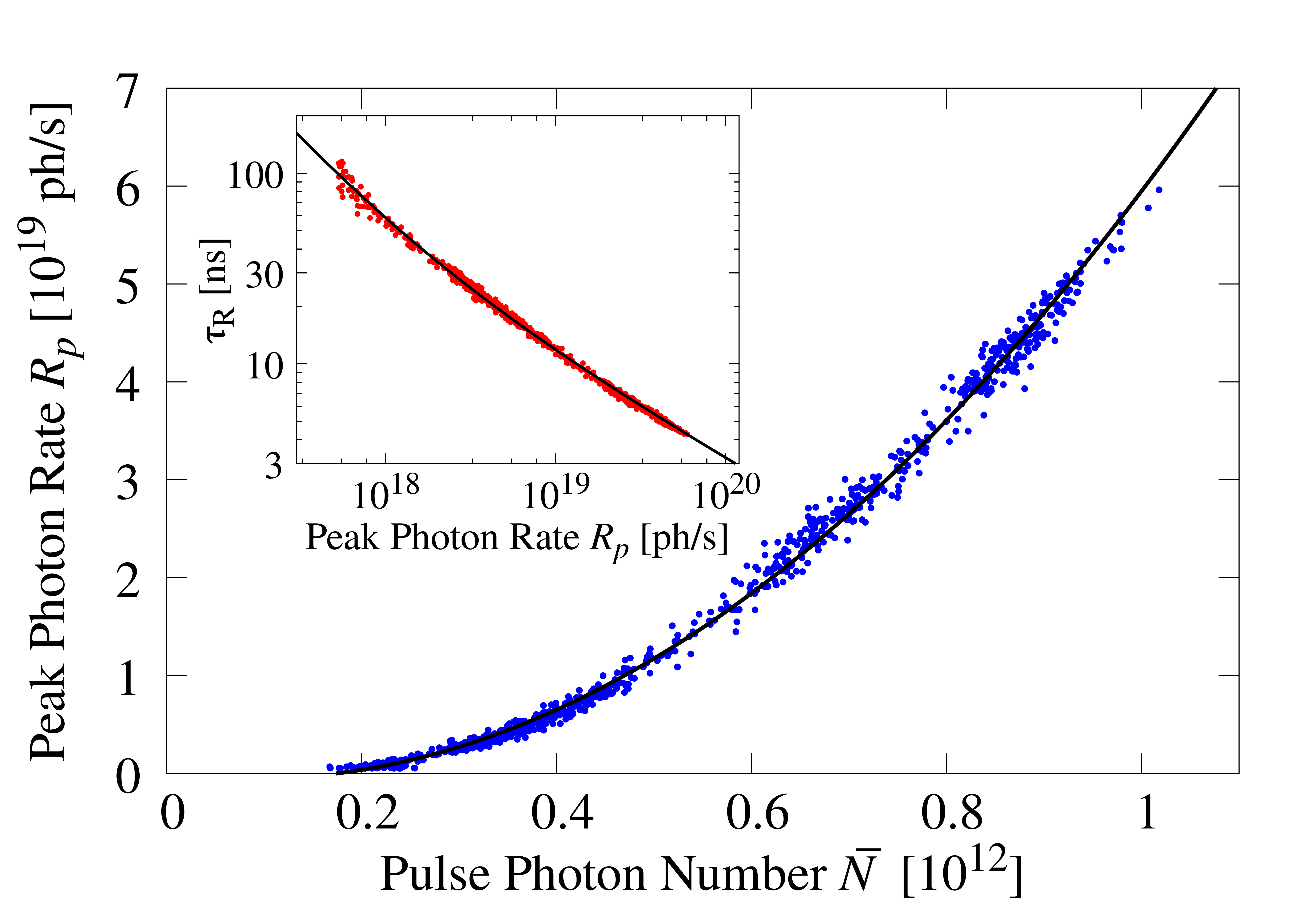}
	\caption{ $R_p$ vs $\bar{N}$ and $\tau_R$ vs $R_p$ (inset) best-fit parameters of the 1.5\,$\mu$m pulses emitted by Er:YLF doped at 1\%. The laser power and beam diameter at the crystal are 400\,mW and 66\,$\mu$m, respectively. The black lines are the results of the fitting procedure with the functions in Eq.\,\ref{eq:Rp1} and Eq.\,\ref{eq:taup1}.}
	\label{Fig:3}
\end{figure}
In fact, $\mu$ and $N_0$ are the only free parameters as $A=14.5\,$s$^{-1}$ is actually estimated independently by measuring both the $^4\mathrm{I}_{13/2}(0)$ lifetime and the related spectral intensity of the 8 radiative transitions towards the ground state manifold. 
For a pencil-shaped, homogeneous atomic sample, the geometry factor is given by $\mu = 3\lambda^2 /(8\pi^2 \omega_0^2)$, where $\omega_0$ is the incident laser beam waist and $\lambda$ is the emission wavelength\,\cite{Rehler1971,Gross1982}. 
For the data acquired with $\omega_0=66\,\mu$m and shown in Fig.\,\ref{Fig:3}, the expected value $\mu=(2.0 \pm 0.2) \cdot 10^{-5}$ agrees with the obtained best-fit parameter $\mu=(1.9 \pm 0.3) \cdot 10^{-5}$. We recorded SF pulses also with $\omega_0=130\,\mu$m and the calculated value in this case exceeds the experimental one by a factor 4, for both crystals concentration. Such a discrepancy may be ascribed to the fact that the pencil-shaped geometry approximation is in this case not suited to fit the excited atomic distribution.

Our findings show that coherence spontaneously develops over more than ${10^{11}}$ Er$^{3+}$ ions. The largest value $N=8\cdot10^{12}$ ions was observed in the Er:YLF 1\% crystal with $\omega_0=130\,\mu$m. As previously reported in Er:YSO\,\cite{Braggio2020}, this atom number is a fraction ($\sim$10$^{-3}$) of the inversion population, suggesting a self-selection process of the excited ions based on the similarity of their transition frequency.
 
An interesting aspect of CSF is the correlation between the paired SF pulses, which allows for investigating the $\tau_d$ statistics even in the cw pumping regime. For an ensemble of $N$ identical, uncorrelated atoms, the delay time is determined by the strength of the quantum fluctuations initiating the SF process. Once the coherence is seeded, the physical system evolution is semi-classical and macroscopic fluctuations of $\tau_d$ are a direct consequence of quantum noise\,\cite{Vrehen1980}. 

Rehler and Eberly\,\cite{Rehler1971} derived the average delay time $\braket{\tau_d}= (\mu A N)^{-1}\ln(\mu N) = \tau_R \ln(\mu N)$, and its standard deviation $\sigma( \tau_d)=1.3\tau_R$, whereas for other authors\,\cite{Polder1979a} $\braket{\tau_d}= 0.25\tau_R [\ln(\sqrt{2\pi N})]^2$ and $\sigma( \tau_d)=2.3\braket{\tau_d}/\ln N$ are expected. It is worth noticing that both models neglect the transverse and light propagation effects, the dephasing processes and inhomogeneous line broadening. For these reasons, experimental findings generally differ from expected values, especially in solid state systems\,\cite{Florian1984,Raino2018}. Clearly, a better agreement is found with optically trapped ions\,\cite{Norcia2016,Laske2019}.

\begin{figure}[]
	\includegraphics[width=0.5\textwidth]{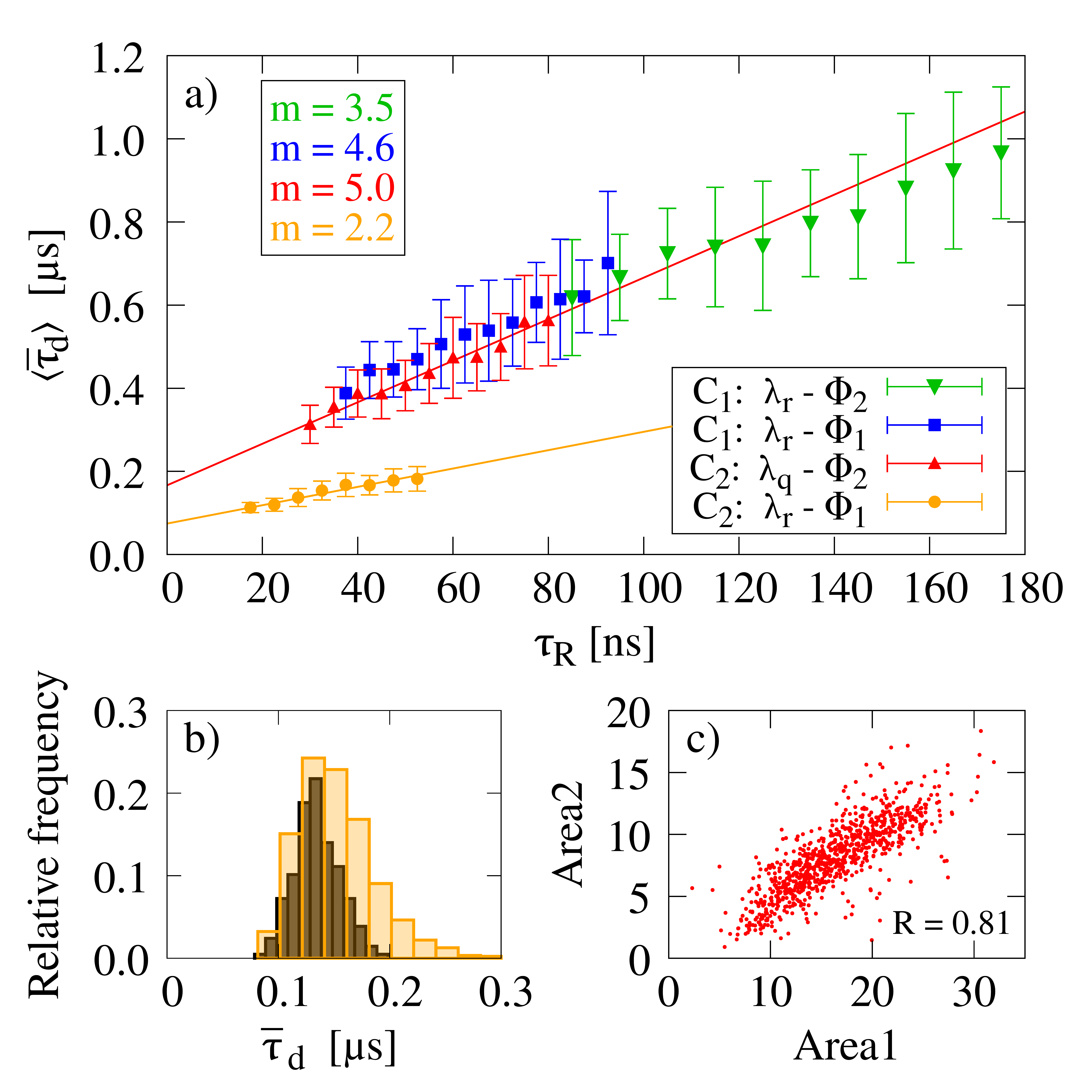}
	\caption{(a) $\tau_R$ vs $\bar{\tau}_d$ for Er:YLF 0.01\% (C1) and Er:YLF 1\% (C2) crystals, with the pump laser tuned at $\lambda_r=808.9921$\,nm or $\lambda_q=808.9964$\,nm, and beam waist of 130\,$\mu$m ($\Phi_1$) or 66\,$\mu$m ($\Phi_2$). For the sake of clarity, only the linear fits corresponding to the Er:YLF 1\% data sets are drawn. (b) Observed probability distribution of the $\bar{\tau}_d$ (orange histogram). The same distribution for 25\,ns\,$<\tau_R<$\,30\,ns is shown in black. (c) Correlation between the area of the first and the second pulse for the red data set. Uncalibrated units are used for both axis. $R$ denotes the correlation coefficient.}
	\label{Fig:4}
\end{figure}

As previoulsy demonstrated for a similar pumping scheme\,\cite{Laske2019}, the relaxation process within the $^4\mathrm{I}_{13/2}$ manifold, that takes place before the 1534\,nm-wavelength SF onset, shifts the distribution peak position to $\braket{\bar{\tau}_d}=\braket{\tau_d}+t_0$.
In Fig.\,\ref{Fig:4}\,a we report $\braket{\bar{\tau}_d}$ versus $\tau_R$ of the second pulse for four different experimental conditions. Each data set is obtained from approximately one thousand recorded paired pulses. The points represent the average $\bar{\tau}_d$ for the binned $\tau_R$ values, with 5\,ns (orange, red and blue data) or 10\,ns (green data) bin widths. The error bars indicate the corresponding standard deviation and hence the amplitude of the delay time fluctuations. An example of the $\bar{\tau}_d$ distribution is shown in Fig.\,\ref{Fig:4}\,b. 

The $\braket{\bar{\tau_d}}$ values linearly depend on $\tau_R$ for all four data sets, as expected for SF. The same trend applies to $\sigma(\tau_d)$ values, indicating larger fluctuations for smaller ensembles. However, the estimated $\braket{\bar{\tau_d}}$-$\tau_R$ linear coefficients $m$, constrained within 2 and 5, are smaller than those calculated for $N=$10$^{11}$-$10^{12}$ using the mentioned simplified models. At resonance, the pump laser absorption is strong in the 1\%-concentration sample and the excited ion are mainly found within the first hundreds of micrometers, as is the case for the orange data for Fig.\,\ref{Fig:4}. These data significantly differ from the other sets in terms of linear coefficient, line intercept and fluctuations amplitude, as the distribution is less uniform along the whole length of the crystal ($\sim5$\,mm). This might be explained by considering radiating transverse modes, whose number equals the square of the Fresnel factor $F=\pi \omega_0^2/\lambda L$, with $L$ the length of the atomic ensemble\,\cite{Mostowski1983a}.  Authors have in fact reported a reduction of both $\braket{\bar{\tau_d}}$ and $\sigma(\tau_d)$ for increasing $F$ values\,\cite{Vrehen1981}. 
The smaller value of the orange data line intercept, which is an estimator for $t_0$, can be also ascribed to the stronger laser absorption that induces a higher phonon density\,\cite{Bottger2006}.

In the CSF of our physical system, a significant correlation is obtained between the area of the paired pulses (see Fig.\,\ref{Fig:4}\,c), despite the influence of dephasing processes and fluctuations of the $^4\mathrm{I}_{13/2}(0)$ steady-state population. 

In this work we demonstrate that it is possible to accomplish CSF in solid-state. The use of cw pumping enables the detection of light pulses stemming from well-identified macrocoherent states, whose temporal dynamics is in good agreement with models that consider the number of radiating ions and their spatial distribution. Most importantly, the cascade superfluorescence allows for investigating also the delay time and its fluctuations in physical systems where SF is obtained with cw pumping.


\section{Author contributions}
F.C. and C.B. designed and performed the experiments with the help of A.K.. F.C. and C.B. wrote the manuscript. F.C. analyzed the data. A.D.L. and M.T. grew and characterized the samples. All authors discussed the results and revised the manuscript. The authors declare that they have no competing interests.

\noindent

 \bibliography{My_Collection}

\end{document}